\documentstyle[epsf]{mn}
\newif\ifAMStwofonts
\AMStwofontstrue

\ifoldfss
  \ifCUPmtlplainloaded \else
    \NewTextAlphabet{textbfit} {cmbxti10} {}
    \NewTextAlphabet{textbfss} {cmssbx10} {}
    \NewMathAlphabet{mathbfit} {cmbxti10} {} 
    \NewMathAlphabet{mathbfss} {cmssbx10} {} 
  \fi
  \ifAMStwofonts
    \ifCUPmtlplainloaded \else
      \NewSymbolFont{upmath} {eurm10}
      \NewSymbolFont{AMSa} {msam10}
      \NewMathSymbol{\upi}     {0}{upmath}{19}
      \NewMathSymbol{\umu}     {0}{upmath}{16}
      \NewMathSymbol{\upartial}{0}{upmath}{40}
      \NewMathSymbol{\leqslant}{3}{AMSa}{36}
      \NewMathSymbol{\geqslant}{3}{AMSa}{3E}

    \fi
  \fi
\fi 

\ifnfssone
  \newmathalphabet{\mathit}
  \addtoversion{normal}{\mathit}{cmr}{m}{it}
  \addtoversion{bold}{\mathit}{cmr}{bx}{it}
  \newmathalphabet{\mathbfit} 
  \addtoversion{normal}{\mathbfit}{cmr}{bx}{it}
  \addtoversion{bold}{\mathbfit}{cmr}{bx}{it}
  \newmathalphabet{\mathbfss} 
  \addtoversion{normal}{\mathbfss}{cmss}{bx}{n}
  \addtoversion{bold}{\mathbfss}{cmss}{bx}{n}
  \ifAMStwofonts
    \ifCUPmtlplainloaded \else
      \UseAMStwoboldmath
      \makeatletter
      \new@mathgroup\upmath@group
      \define@mathgroup\mv@normal\upmath@group{eur}{m}{n}
      \define@mathgroup\mv@bold\upmath@group{eur}{b}{n}
      \edef\UPM{\hexnumber\upmath@group}
      \new@mathgroup\amsa@group
      \define@mathgroup\mv@normal\amsa@group{msa}{m}{n}
      \define@mathgroup\mv@bold\amsa@group{msa}{m}{n}
      \edef\AMSa{\hexnumber\amsa@group}
      \makeatother
      \mathchardef\upi="0\UPM19
      \mathchardef\umu="0\UPM16
      \mathchardef\upartial="0\UPM40
      \mathchardef\leqslant="3\AMSa36
      \mathchardef\geqslant="3\AMSa3E
    \fi
  \fi
\fi 

\ifnfsstwo
  \DeclareMathAlphabet{\mathbfit}{OT1}{cmr}{bx}{it}
  \SetMathAlphabet\mathbfit{bold}{OT1}{cmr}{bx}{it}
  \DeclareMathAlphabet{\mathbfss}{OT1}{cmss}{bx}{n}
  \SetMathAlphabet\mathbfss{bold}{OT1}{cmss}{bx}{n}
  \ifAMStwofonts
    \ifCUPmtlplainloaded \else
      \DeclareSymbolFont{UPM}{U}{eur}{m}{n}
      \SetSymbolFont{UPM}{bold}{U}{eur}{b}{n}
      \DeclareSymbolFont{AMSa}{U}{msa}{m}{n}
      \DeclareMathSymbol{\upi}{0}{UPM}{"19}
      \DeclareMathSymbol{\umu}{0}{UPM}{"16}
      \DeclareMathSymbol{\upartial}{0}{UPM}{"40}
      \DeclareMathSymbol{\leqslant}{3}{AMSa}{"36}
      \DeclareMathSymbol{\geqslant}{3}{AMSa}{"3E}
    \fi
  \fi
\fi 

\ifCUPmtlplainloaded \else
  \ifAMStwofonts \else 
    \def\upi{\pi}
    \def\umu{\mu}
    \def\upartial{\partial}
  \fi
\fi

\title{Isotropic Wavelets: a Powerful Tool to Extract
Point Sources from CMB Maps}
\author[] {L. Cay\'on$^{1}$, J.L. Sanz$^{1,2}$, R.B. Barreiro$^{1,3,4}$,
E. Mart\'\i nez-Gonz\'alez$^{1}$, P. Vielva$^{1,3}$,\\ 
\newauthor
L. Toffolatti$^{5}$, J. Silk$^{6}$, J.M. Diego$^{1,3}$ \& F. Arg\"ueso$^{7}$\\
1. Instituto de F\'\i sica de Cantabria, Fac. Ciencias, Av. los
	Castros s/n, 39005 Santander, Spain\\
2. On sabbatical leave at Center for Particle Astrophysics and Astronomy Department, University of California, Berkeley, USA\\
3. Departamento de F\'\i sica Moderna, Universidad de Cantabria, 
	39005 Santander, Spain.\\
4. Astrophysics Group, Cavendish Laboratory, Madingley Road, Cambridge CB3 0HE, UK\\
5. Dpto. de F\'\i sica, Universidad de Oviedo, c/ Calvo Sotelo s/n, 33007 Oviedo, Spain\\
6. Astronomy Department, University of Oxford, UK\\
7. Dpto. de Matem\'aticas, Universidad de Oviedo, c/ Calvo Sotelo s/n, 33007 Oviedo, Spain\\}
\date{\today}

\pagerange{\pageref{firstpage}--\pageref{lastpage}}
\pubyear{1998}

\begin{document}

\maketitle

\label{firstpage}

\begin{abstract}
It is the aim of this paper to introduce the use of isotropic
wavelets to detect and determine the flux of point sources
appearing in CMB maps. 
The most suited wavelet to detect point sources
filtered with a Gaussian beam is the Mexican Hat. An analytical expression
of the wavelet coefficient obtained in the presence of a point source
is provided and used in the detection and flux estimation 
methods presented. For illustration the method is applied to 
two simulations (assuming Planck Mission characteristics) dominated
by CMB (100 GHz) and dust (857 GHz) as these will be the two signals
dominating at low and high frequency respectively in the
Planck channels.
We are able to detect bright sources above
$1.58$ Jy at 857 GHz ($82\%$ of all sources) and above $0.36$ Jy at 100 GHz ($100\%$ 
of all) 
with errors in the flux
estimation below $25\%$. The main advantage of this method is that 
nothing has to be assumed about the underlying field, i.e. about
the nature and properties of the signal plus noise present in the maps.
This is not the case in the detection method 
presented by Tegmark and Oliveira-Costa 1998. Both methods are
compared producing similar results.

\end{abstract}


\section{Introduction}

	One of the major challenges in the study of the Cosmic Microwave 
Background (CMB) at present, is to overcome the problem of
separating the cosmological signal from the Galactic foreground emissions,
the noise and the extragalactic point sources, in the future
high resolution CMB maps. Such missions as MAP (Bennett et al. 1996) 
and Planck (Mandolesi et al. 1998; Puget et al. 1998) will produce
CMB maps that will be seriously affected by this problem. 

Several methods have already been tested to perform the separation of the
different emissions in CMB observations, as the ones based on
the Wiener filter (Tegmark and Efstathiou 1996, Bouchet and Gispert 1999) 
and on Maximum Entropy 
Methods (Hobson et al. 1998, 1999). Recently, wavelet techniques 
have been introduced in the analysis of CMB data. Denoising of CMB maps 
has been performed on patches of the sky of $12^{\circ }.8\times 12^{\circ}.8$
using multiresolution techniques (Sanz et al. 1999a) 
and 2D wavelets (Sanz et al. 1999b), as well as on the whole celestial 
sphere (Tenorio et al. 1999).

Extragalactic
point sources should be removed from the maps before any analysis is 
performed. There are still many uncertainties regarding the 
composition of this population, the characteristics
of the sources, and their time variability as well as the abundance 
and emission at different frequencies. 
Simulations of the emission of extragalactic sources at the
observing frequencies of the Planck mission have been worked out by
Toffolatti et al. 1998 and Guiderdoni 1999. In general there is 
agreement between these extrapolations. In this paper we use simulations
based on the predictions of Toffolatti et al. 1998. The number counts 
of  bright objects at frequencies above 300 GHz are dominated by
far-IR selected sources. Lower frequencies will get most of the emission 
from radio selected sources.

Removal of point sources in the observed maps is usually done by
removing all pixels above a threshold of $5\sigma $. The error in 
rejecting pixels not corresponding to point sources is very small. However,
large number of point sources will still remain undetected.
Tegmark and Oliveira-Costa 1998 (TOC98 hereinafter) presented a 
method to locate and detect
point sources based on the minimization of the variance of the map. 
The application of this technique assumes that the
CMB, noise and Galactic foregrounds behave as Gaussian fields and 
it requires knowledge of the power spectrum of  these components.
As a product of the MEM used to separate all the foreground components
from the CMB, Hobson et al. 1999 are able to recover 
the point sources present in the simulated maps included as part
of the noise in the method. However they are not able 
to remove the brightest sources that can still be 
observed in the residuals. As will be shown these will be the point 
sources that
the method presented here is able to extract.
Tenorio et al. 1999 applied Daubechies wavelets to detect and remove
point sources from CMB plus noise flat maps.
However Daubechies wavelets are not the optimal
ones to detect point sources as explained below.
As part of the effort to explore the possibilities of using wavelets
to extract the information contained in CMB observations, we are 
studying in this paper isotropic wavelets in the context of this data.
In particular, taking advantage of the spherical symmetry, these wavelets
are very appropiate for locating and detecting point sources. 
The method does 
not assume any behaviour for the rest of the signals and noise present
in the maps. The basic point source characteristics are its Gaussian 
shape and scale,  defined by the beam used in the observations. 
Convolution of the CMB map with a wavelet of the same 
scale and similar shape of the point source will produce wavelet coefficients 
with maxima (amplification) at the position of point sources (location). 
The most appropiate wavelet to detect Gaussian shaped sources is the
so called Mexican Hat. This wavelet has been used to detect sources 
in the presence of noise in X-ray images by Damiani et al. 1997.

The paper is organized as follows. An analytical approach to the use
of isotropic wavelets to analyse CMB maps is presented in section 2.
Section 3 is dedicated to the application of isotropic wavelets to 
CMB maps in order to detect point sources. A comparison
with the TOC98 method is also included in this section. We present the flux
estimation of the detected point sources in section 4. Wherever flux 
magnitudes are given they refer to the total integrated flux under the beam. 
Conclusions are included in section 5.

\section{Analytical Approach}

The continous isotropic wavelet transform of a 2D signal
$f(\vec{x})$ is defined as

\begin{eqnarray}
wv(R, \vec{b}) = \int d\vec{x}\,f(\vec{x})\Psi (R, \vec{b}; \vec{x})\; ,\nonumber
\\
\Psi (R, \vec{b}; \vec{x}) = \frac{1}{R}\psi (\frac{|\vec{x} - \vec{b}|}{R})\; ,
\end{eqnarray}

\noindent where $wv(R, \vec{b})$ is the wavelet coefficient associated with
 the scale 
$R$ at the point with coordinates $\vec{b}$. $\psi (|\vec{x}|)$ is the 
``mother'' wavelet that is assumed to be isotropic and satisfies the constraint
$\int d\vec{x}\,\psi = 0,\int d\vec{x}\,{\psi}^2 = 1$ and the admissibility 
condition: $C_{\psi} = (2\pi )^2\int_0^{\infty}dk\,k^{-1}{\psi}^2(k) < \infty $, where 
$\psi (k)$ is 
the Fourier transform of $\psi (x)$. This is a necessary and sufficient condition 
in order to synthesise  the function with the wavelet coefficients

\begin{equation}
f(\vec{x}) = \frac{1}{C_{\psi}}\int dR\,d\vec{b}\,\frac{1}{R^4}wv(R, \vec{b})
\psi (\frac{|\vec{x} - \vec{b}|}{R})\; .
\end{equation}

Our aim is to locate point sources in microwave maps. The wavelet
coefficients given by equation (1), provide information about the 
contribution of different scales ($R$) to the value of the analyzed 
function ($f$) at
a certain location ($\vec b$). Moreover, they will increase as the 
shape of the wavelet ($\Psi$) gets closer to the shape of the analyzed 
function. We want to maximize the wavelet coefficients at the 
location of point sources. As observations are performed through 
antennas, the observed point sources will be the result of a convolution 
with a certain antenna response function. The response is most frequently 
modeled as a Gaussian. The most appropiate isotropic wavelet to use
in this case is the so called Mexican Hat wavelet given by:  

\begin{equation}
\psi (x) = {(2\pi )}^{-1/2}(2 - x^2)e^{-x^2/2}\; .
\end{equation}

\noindent The convolution of the Mexican Hat wavelet with an image
consistent of  point sources filtered with a Gaussian beam, 
is a functional proportional to the
second derivative of the analyzed function and therefore 
related to the maxima and minima in the analyzed image. 
Moreover the wavelet coefficients corresponding to a constant background 
will be identically zero. 
We would also 
like to point out the resemblance between the Mexican Hat wavelet
and the filter used in TOC98 who  obtained  the optimal filter
to detect point sources.

Observations of the CMB contain not only the cosmological CMB signal
but also Galactic emission (synchrotron, free-free and dust), noise
and extragalactic emission (Sunyaev-Zeldovich effect and point sources).
We are interested in removing point sources (some of the Sunyaev-Zeldovich
sources will be extended and we will deal with the removal
of these in a future paper) from the underlying 
signal (CMB or Galactic emissions) plus noise image.
If one assumes that the sources that will appear in real maps at different 
frequencies correspond to the convolution of a point
source with a Gaussian of dispersion ${\sigma}_a$ 
(antenna beamwidth), then any source (we are only concerned with the
detection of point sources in this paper and will therefore
use the word source to refer to the filtered point sources unless
otherwise indicated) can be represented by a function:
$f(\vec{x}) = (B/A)e^{-{(\vec{x} - \vec{x_o})}^2/2{\sigma}_a^2}$, $B$ being 
the flux of the point source and $A$ the area under the beam. 
At the position of the peak of the source one has the following 
wavelet coefficient at scale $R$:

\begin{equation}
\frac{wv(R, \vec{x_o})}{R} = 2{(2\pi )}^{1/2}(B/A)\,x
{(1 + x)}^{-2}\; ,\ \ \ x\equiv {(\frac{R}{{\sigma}_a})}^2\; ,
\end{equation}

\noindent whereas the wavelet variance of the signal plus noise at scale $R$ 
is given by the expression 

\begin{equation}
\frac{\sigma_{wv}^2 (R)}{R^2} = 4\pi {\sigma}_s^2
\biggl[ \frac{(\sigma_p/R)^2}{SNR^2}+
\frac{\Gamma(\frac{n+7}{2})}{2\Gamma(\frac{n+3}{2})}
\frac{x^2}{{(1 + x)}^{\frac{n+7}{2}}}\biggr]\; ,
\end{equation}

\noindent where $\sigma_s,\sigma_p$ are respectively 
the signal dispersion in real
space and the pixel scale $\sigma_p=l_p/2\pi$ ($l_p$ being the pixel
size), and it is assumed that $R>>\sigma_p$. $SNR$ is the signal to noise
ratio of the signal plus noise map.
The signal has been represented by a power spectrum with effective index $n$. 
Consider the amplification suffered by the flux of a source
as the ratio of the detection level in wavelet space $(wv(R,\vec x_o)/
\sigma_{wv}(R))$ to the detection level
in real space $((B/A)/\sigma_{s+n})$. This quantity will
be maximum at $R\simeq \sigma_a$ as  
the
wavelet at that scale selects the contribution of scales around the size 
of the source whereas the 
signal and noise appear at a higher or lower scale, respectively. In
theory,
one can calculate the amplification taking into account equations (4) and
(5). The amplification factor is obtained to be greater than 2.3 for 
CMB dominated maps and larger in the case of dust dominated maps. This
result is also obtained numerically as will be shown in the next section.

\setcounter{figure}{0}
\begin{figure}
 \epsfxsize=3in
 \epsffile{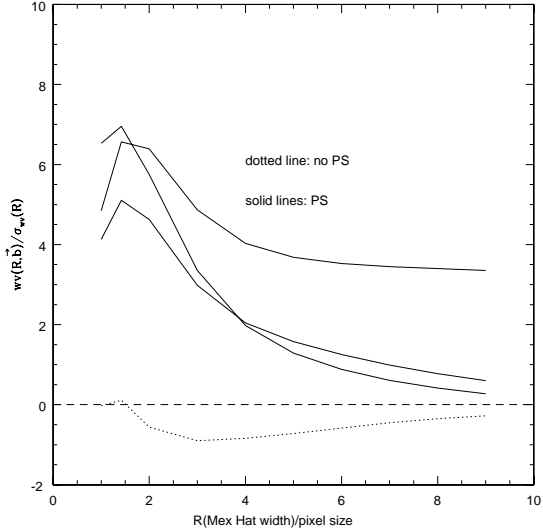}
 \caption{Wavelet coefficients normalized to the dispersion of the
wavelet coefficient map as a function of the Mexican Hat width $R$ in
units of number of pixels. Each of the curves corresponds to different
locations in the analyzed map (different pixels in the original map
that includes signal (dust), noise and point sources). Pixels occupied by a 
point source are represented by solid lines. The dotted line 
corresponds to a pixel where there is no point source.}
 \label{f1}
\end{figure}

\section{Point Source Detection}

Isotropic wavelets are 
defined as a function of 
position $\vec b$ and scale $R$. As pointed out in the previous section, 
the optimal wavelet scale in order to detect point sources in CMB maps
is approximately the dispersion of the antenna $\simeq \sigma_{a}$. 
Convolution with a Mexican Hat of that scale will produce peaks at the 
positions of point sources and will allow for point source 
detection.

Our aim is to illustrate the performance of the detection technique
considering representative maps of high sensitivity experiments. 
We have chosen to simulate maps following the Planck Mission
characteristics.
The detection method is applied to simulations of flat patches 
of the sky of $12^{\circ}.8\times 12^{\circ}.8$
at two different frequencies, 100 and 857 GHz, 
with angular resolution (arcmin)=10 (100 GHz), 5 (857 GHz) and
pixel size (arcmin)=3 (100 GHz), 1.5 (857 GHz). As previously pointed out,
CMB observations will contain the cosmological signal, Galactic
foregrounds emissions, extragalactic 
emissions and instrumental noise. In the frequency range that the
Planck Mission will sample, the observed maps
will be dominated either by the cosmological CMB signal at
low frequencies (below $\sim 200$ GHz) or by dust at high frequencies.
As two representative extreme cases we have chosen $100$ and $857$ GHz
to present the detection method based on isotropic wavelets.  
Therefore simulations at 100 GHz include cosmological signal, point sources 
and instrumental noise and simulations at 857 GHz consist of dust emission, 
point sources and instrumental noise. Cosmological signal simulations 
are performed of Cold Dark Matter (CDM) model with $\Omega =1$. Dust
simulations are made using the predicted full-sky maps of
Finkbeiner et. al. 1999. These predictions assume that dust emission
corresponds to two grey bodies, with an emissivity of 1.67 for the cold
one and 2.70 for the hot one (this has been checked in the frequency
range of FIRAS). Point source simulations
at the two frequencies considered are based on
the code developed by Toffolatti et al. 1998. Non correlated
Gaussian noise is added to the signal with a $\Delta T/T$ per
resolution element of $4.3\times 10^{-6}$ at 100 GHz and
$6670\times 10^{-6}$ at 857 GHz as in the corresponding two channels 
of the Planck mission.

To show how different Mexican Hat widths affect the value of the 
wavelet coefficients at positions in the simulated map with or
without a point source, we have plotted in Figure 1 the value 
of the wavelet coefficient normalized to the dispersion of the
wavelet coefficient map as a function of the Mexican Hat width $R$. It is
clear that pixels with a point source have a characteristic 
curve with a maximum at $R\simeq \sigma_a$. Taking this 
behaviour into account, the  
detection method proceeds in three steps. First the simulated map is convolved with Mexican Hat wavelets of several widths (R) obtaining maps of wavelet
coefficients at each R. In the second step we look at the map of wavelet
coefficients obtained for $R=\sigma_a$ (scale at which the 
amplification is maximum) and find the spots (connected
pixels) appearing above $5\sigma_{wv}(R=\sigma_a)$. The location of the maxima
of these spots will correspond to the location of point sources as previously 
discussed. In order to make sure that all detections correspond
to real point sources
we determine, in the third step, the numerical curve of 
$wv(R,\vec x_o)/\sigma_{wv}(R)$ as 
a function of $R$ (being $\vec x_o$ the position of a maximum identified 
in the second step) and compare it with the theoretical one given
by equations (4) and (5). A reduced $\chi^2$ is calculated and only maxima
positions with values of  $\chi^2\approx 1$ are accepted as point source 
detections. 

We have plotted in Figure 2 the number of point sources detected above a 
certain flux together with the number of point sources that 
exist in the simulation after filtering (notice that the number of 
point sources above a certain flux can change when a Gaussian filter is 
applied to the simulated map, in particular 
at high frequencies and low fluxes). 
As one can see the detected point sources are the ones with the 
highest fluxes. The method is best suited to detect the bright
sources leaving the majority of the faint ones undetected.
At $857$ GHz the number of point sources 
detected using the wavelet method is $82\%$ of the ones that
exist in the simulation with fluxes above $1.58$ Jy. At 100 GHz we are
able to detect $100\%$ of all the point sources existing above $0.36$ Jy.
We have also looked at the amplification suffered by the detected
point sources (once the point source is located we compare the 
amplitude of that pixel in the wavelet coefficients map relative to the
dispersion of that map with the amplitude of that pixel in the
original (signal plus noise plus point sources) map relative to the
dispersion of that map) obtaining average values of 
11 and 4 at 857 GHz and 100 GHz respectively.

\setcounter{figure}{1}
\begin{figure}
 \epsfxsize=3in
 \epsffile{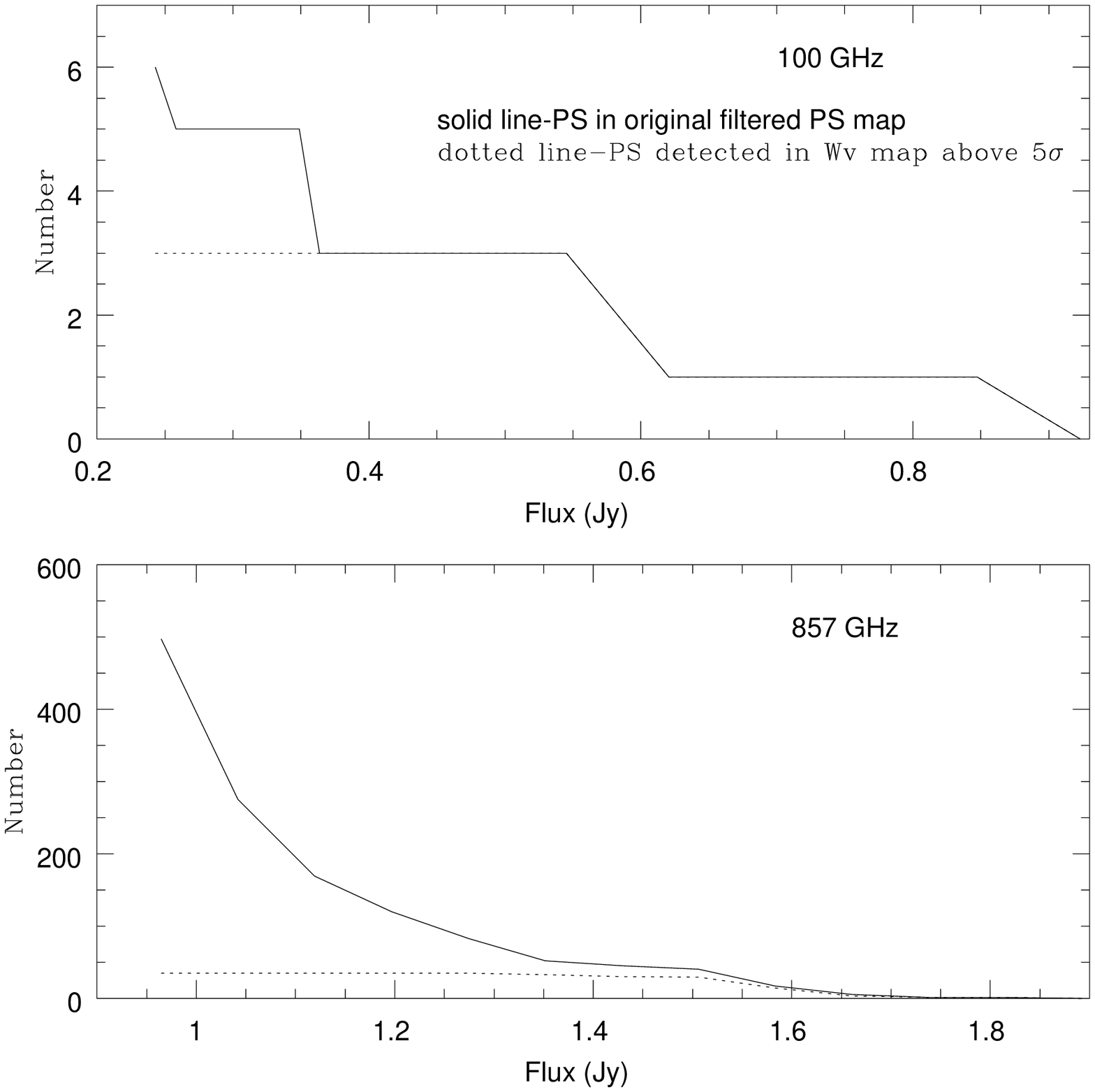}
 \caption{In each of the two pannels the dotted line 
shows the number of point sources
detected by the Wavelet method above a given threshold. The number of simulatedsources after filtering with a Gaussian beam of $FWHM=10',5'$ at 
100 and 857 GHz respectively, 
is indicated by a solid line. 
Top and bottom figures correspond
to the simulations at 100 and 857 GHz respectively.
}
 \label{f2}
\end{figure}

We have compared our results with the method used in TOC98. 
To illustrate how both methods perform on the 
same data we have considered the same simulations at
100 GHz and 857 GHz. The number of point sources detected by the
two methods are given in Table 1 (also included in the second
column is the number of point sources directly found above $5\sigma$ in the
signal plus noise plus point sources map). 
As one can see, the performance of both methods is comparable. 
Moreover the average amplification suffered by the detected 
point sources is also comparable. 
The detection is clearly
improved in relation to considering only peaks above $5\sigma $. 
This improvement is a consequence of the amplification suffered by
the amplitude of the point sources after applying the Mexican Hat 
wavelet (with the appropiate scale).
It is important to note that no 
assumptions about the nature of the underlying signal plus noise
field are made in the wavelet method presented in this paper.
However, the detection method of TOC98 is based on modeling 
the signal plus noise present in the analyzed maps as Gaussian
random fields and requires  a knowledge of the power
spectrum of all of the signal plus the noise components.

\begin{table}
  \begin{center}
  \caption[]{Number of point sources detected above $5\sigma$ and using the wavelet and the TOC98 methods}
  \label{tab1}
  \begin{tabular}{c|c|c|c}\hline
  Frequency (GHz) & Above $5\sigma$ & Wavelet & TOC98 \cr
  \hline\hline

	857 & 8 & 35 & 32\cr
	100 & 0 & 3 &  3\cr

  \hline\hline
  \end{tabular}
  \end{center}
\end{table}

\section{Flux Estimation of the Detected Point Sources}

In this section we want to show how the flux of the 
detected point sources can be recovered using wavelet
coefficients. From equation (4) we know the analytical expression of the 
wavelet coefficient at the position of the detected point source, being
proportional to the filtered flux of the point source $B$. 
Moreover, to take into account the information at different scales
we obtain the final amplitude value from the $\chi^2$ fit presented in
section 3.
Defining the 
error in the flux estimation as the ratio of the
difference between the real and the recovered
value divided by the real value, we have plotted in Figure 3 these
errors for the point sources detected in the two simulated maps. At 
857 GHz point sources with fluxes above $1.58$ Jy have
errors with absolute values below $20\%$, 
and $46\%$ of all detections have errors with absolute values below $10\%$.
The three point sources detected at 100 GHz have errors with
absolute values below $25\%$.
As one can see from the figure the overestimated amplitudes have 
larger errors than the underestimated ones (this can only
be seeing in the 857 GHz results as the statistics of detected
point sources is very poor at 100 GHz). There is also a slight
bias towards negative error values. 
The $\chi^2$ fit used to recover the amplitude does not take into
account possible correlations between the different scales. Including
these correlations improves 
the amplitude determination method so that
no bias appears. This will be presented in a future paper.

\setcounter{figure}{2}
\begin{figure}
 \epsfxsize=3in
 \epsffile{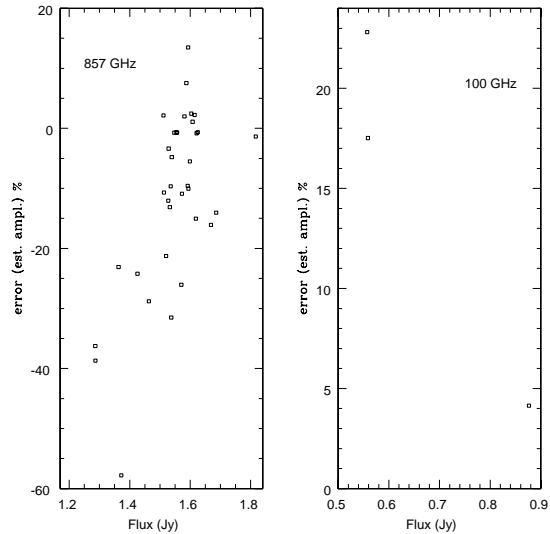}
 \caption{Error in the estimated flux of the point sources
detected at 857 and 100 GHz (left and right panels respectively)
as a function of the input flux.  
}
 \label{f3}
\end{figure}

Once the position and the flux of the detected point sources
is known, one can simulate a map with only point sources filtered with
the corresponding Gaussian beam (depending on the frequency) and 
subtract it from the original map. A map with point sources removed
will be obtained. This will be an improvement on  the map obtained 
by setting to 0 the value of the pixels affected by the detected point sources
as is usually done.

\section{Conclusions}

Wavelet coefficients provide information at each location of an 
analyzed image of the contributions from the different scales. Moreover,
as they are obtained from the convolution of the image with the wavelet,
in the case that the shape of the wavelet coincides with the shape 
of the field at a fixed position, the wavelet coefficient will be
maximum. Since we are interested in extracting point sources from
CMB maps, isotropic wavelets have the appropiate shape. 
CMB observations are performed using antennas
with a characteristic response. Modeling the antenna as a Gaussian beam
the observed point sources will be the result of convolving with 
such a beam. In this case, the best suited 
isotropic wavelet to detect point sources 
is the Mexican Hat. This wavelet has
a very similar shape to the optimal filter used by TOC98. 

As a demonstration of the method, the study is performed 
on simulations at 100 GHz and 857 GHz dominated by CMB and dust
respectively and with the characteristics of the future Planck Mission.
We have used the Mexican Hat wavelet to detect and determine the
flux of point sources in $12.8\times12.8$ square degree 
simulations. 
The detection method is based on the maxima reached by the wavelet
coefficients at $R\simeq \sigma_a$ at point source positions. 
The reason for this is that the 
signal and the noise have characteristic scales larger and smaller 
than $\sigma_a$ respectively
and the wavelet used has the appropiate shape to resemble 
that of a point source filtered with a Gaussian. 
We 
consider detections above 5 $\sigma_{wv}(R=\sigma_a)$. Making use of 
the information provided by the wavelet coefficients obtained 
convolving with Mexican Hat wavelets of different widths (R) as well
as of the analytical expression we make sure that all detections 
correspond to real point sources by a reduced $\chi^2$ fit. This method
detects point sources in the
simulated maps above $\approx 1.2$ Jy at 857 GHz and above $\approx 0.5$ Jy
at 100 GHz. $82\%$ of the existing point sources above $1.58$ Jy are
detected in the simulation at 857 GHz whereas all the existing point sources 
above $0.36$ Jy are detected at 100 GHz. The detections are achieved
due to the amplification suffered by the ratio of 
the flux of the point source 
in the wavelet coefficient map (at $R=\sigma_a$) relative to the
dispersion of that map in comparison
with the flux in the original simulated map relative to its dispersion. 
Average 
amplification values of 11 and 4 are found at 857 GHz and 100 GHz. 
It is very important to note  that the detection method is not based on any
assumption about the nature of the signal plus noise. The method
presented in TOC98 provides comparable
results as presented in section 3. 
TOC98 method uses the optimal filter to locate point sources
in CMB maps however a price has to be paid: the signal and noise
are assumed to be Gaussian random fields and a knowledge of the power
spectrum is required. 
We would also like to point out the possibility of using this method as
complementary to the MEM presented in Hobson et al. 1999. The bright
point sources could be removed from the maps
before MEM is applied. The 
rest of the point sources still in the analyzed maps will be recovered
as a product of the separation method as explained in their paper.

Finally, based on the analytical knowledge of the wavelet coefficients
at positions dominated by point sources we recover the flux
of the detected point sources. For the extracted point sources
with fluxes above $1.58$ Jy at 857 GHz the 
errors (absolute value) are below $20\%$.
The flux of the point sources detected in the simulation at 100 GHz
have errors (absolute value) below $25\%$. Removal of these point sources from
the original map can therefore be done by simulating maps of only point sources
at the locations found in the detection procedure and subtracting
them from the initial ones. We would also like to point out that an analysis
of the whole sphere could also be performed by just dividing the sphere 
in small flat patches.

This paper is aimed to serve as a presentation of the power of
isotropic wavelets to detect sources in CMB maps. No information
on different frequencies has been used in this work. In a future paper
we will add this information and will study the separation of 
Sunyaev-Zeldovich sources from point sources in Planck simulated maps.

\section*{Acknowledgments}
This work has been supported by the Comisi\'on Conjunta Hispano-Norteamericana
de Cooperaci\'on Cient\'\i fica y Tecnol\'ogica ref. 98138, 
Spanish DGES Project
no. PB95-1132-C02-02, Spanish CICYT Acci{\'o}n Especial no. ESP98-1545-E and
the EEC project INTAS-OPEN-97-1992. 
L.C. thanks the CfPA hospitality during May 1999.
J.L.S. acknowledges partial financial support from a Spanish MEC fellowship
and CfPA and thanks Berkeley Astronomy Dept. hospitality during 
year 1999.
R.B.B. thanks finantial support from Spanish MEC in the form of a predoctoral 
fellowship and from the PPARC in the form of a research grant.
PV acknowledges support from an Universidad de Cantabria fellowship.
LT acknowledges partial financial support from the Italian ASI and CNR. 
JMD acknowledges support from a Spanish MEC fellowship FP96 20194004.

\end{document}